\documentclass[journal=jacsat,manuscript=article]{achemso}

\usepackage{amssymb}
\usepackage{chemformula} % Formula subscripts using \ch{}
\usepackage{mciteplus}
\mciteErrorOnUnknownfalse

\author{Guang Chen}
\email{guang.chen@uconn.edu}
\affiliation[IMS]
{Institute of Materials Science, University of Connecticut, Storrs CT, 06269}

\title{DMInet: an accurate and highly flexible deep learning framework for drug membrane interaction with membrane selectivity} 


\begin{document}


\begin{abstract}
Drug membrane interaction is a very significant bioprocess to consider in drug discovery. Here, we propose a novel deep learning framework coined DMInet to study drug-membrane interactions that leverages large-scale Martini coarse-grained molecular simulations of permeation of drug-like molecules across six different lipid membranes. The network of DMInet receives three inputs, \textit{viz.}, the drug-like molecule, membrane type and spatial distance across membrane thickness, and predicts the potential of mean force with structural resolution across the lipid membrane and membrane selectivity. Inheriting from coarse-grained Martini representation of organic molecules and combined with deep learning, DMInet has the potential for more accelerated high throughput screening in drug discovery across a much larger chemical space than that can be explored by physics-based simulations alone. Moreover, DMInet is highly flexible in its nature and holds the possibilities for other properties prediction without significant change of the architecture. Last but not least, the architecture of DMInet is general and can be applied to other membrane problems involving permeation and selection.
\end{abstract}

\section{Introduction}
Drug membrane interactions (DMI), a transmembrane activity happening in a soft interface between blood and mammalian membrane, is a necessary process to consider for treating diseased cells in drug delivery and many other therapeutic approaches \cite{seydel2009drug,lucio2010drug}. As such, study of passive permeation of drug-like molecules across phospholipid membranes has not only fundamental significance in physical chemistry or biochemistry, but also very practical meanings in pharmaceutical engineering. 

DMI renders an important bioprocess to consider in drug discovery, for example, in the design of effective antimicrobials for bacterial persisters which cause persistent infections \cite{fisher2017persistent,hart2019small}. It has been reported that these pathogens result in around 2.8 million infections and even 35,000 death in the United States each year \cite{avesar2017rapid}. Different from the majority of bacteria which show significant biosynthetic process, bacterial persisters are dormant, making the common antibiotics targeting biosynthetic processes intractable \cite{hurdle2011targeting}. The membrane-targeting agents are promising yet underexploited for treating persisters, which can disrupt the integrity and the functionality of bacterial membranes, and ultimately kill the persisters \cite{hurdle2011targeting,kim2018new}. A crucial need associated with antibiotic design is to fight against multidrug resistance evolved in pathogens during bacteria-antibiotics competition. Naturally occurring antimicrobial peptides (AMPs) is considered as a promising technique for this treatment, which has been widely studied \cite{epand1999diversity,tossi2000amphipathic,zhao2013engineering}. The idea is also to disrupt the membrane functionality of microbial pathogens. Therefore, an important property for these potential agents is the membrane selectivity. That is, it should target bacterial membrane bilayer only, while selectively avoiding mammalian membranes. As a result, a systematic study of potential drugs across a big chemical space interacting with multi-membrane is of great values, which can bring immense information and insights to DMI, and more importantly, boost discovery of potent drugs.

However, a detailed and thorough understanding of DMI is still lacking \cite{menichetti2019drug}. Scientific challenges associated with this problem come from the intrinsic features of drugs/membranes and the existing study methods. For example, how the concentration of the drug, and membrane types and compositions affect the interactions between them and further affect the membrane selectivity. For potential drugs, the complexity of the possible compositions results in a huge yet discrete chemical space, which makes existing study methods, either experiments or computations, intractable as most optimization methods are gradient-based. Experiments guided by theory and computational simulations have been the mainstreams in drug discovery, though both methods have unavoidable limitations, such as time- and cost-consuming. 

%%%%%%%%%%%%%%%%%%%%%%%%%%%%
\begin{figure*}[h]
\centering
\includegraphics[width=0.8\textwidth]{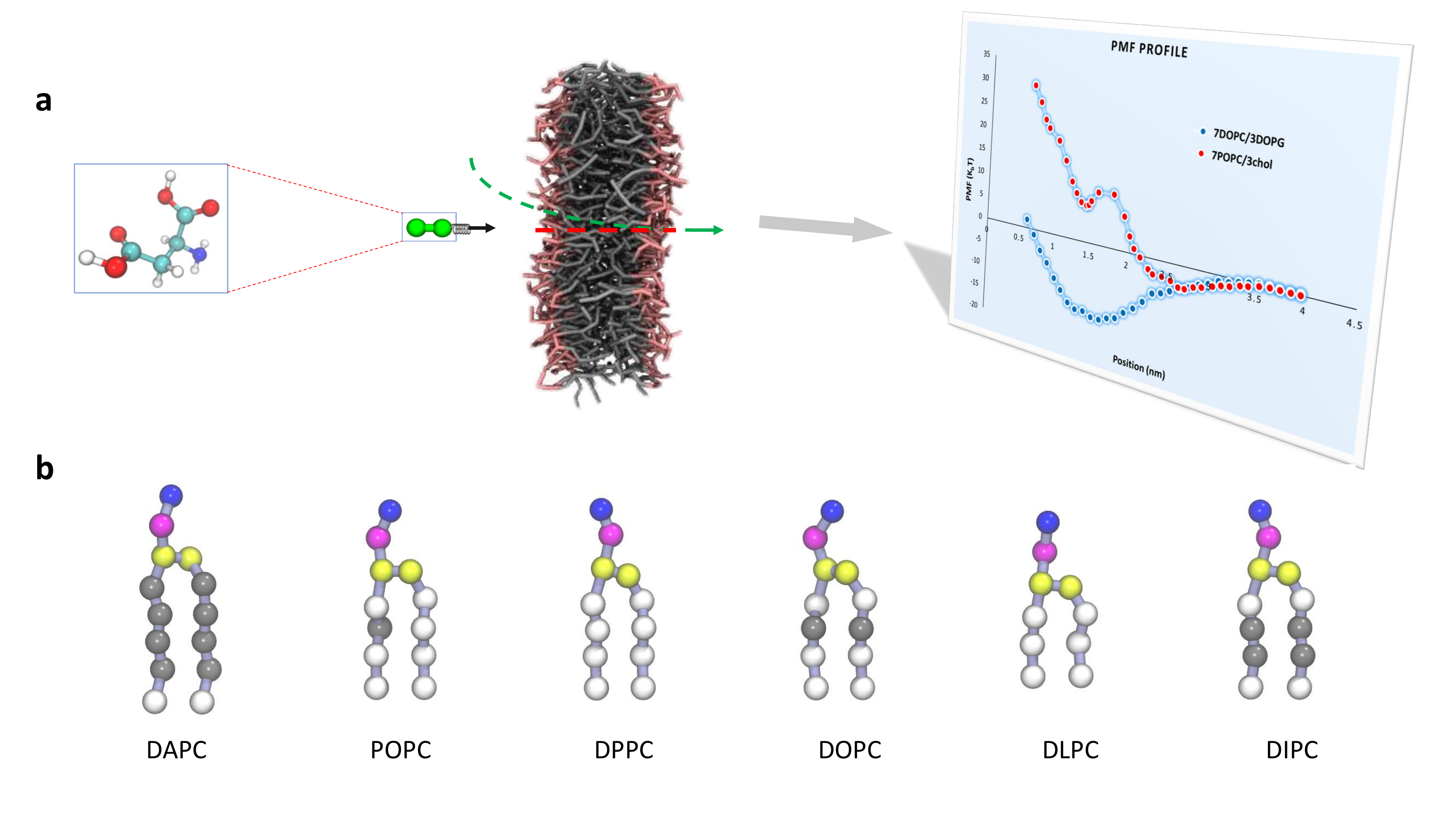}
\caption{Schematic of physical modeling of drug-membrane interaction. \textbf{a}: CGMD simulation of the free energy profile of a sample molecule using umbrella sampling (the PMF profile is drawn after data from Kim \textit{et al.} \cite{kim2019selective}); \textbf{b}: the lipid molecules and corresponding Martini coarse-grained representation (adapted from Ref. \cite{MartiniWeb}).}
\label{Fig:DMI}
\end{figure*}

Despite this limitation, high throughput computation (HTC) for screening has been deemed as the most applicable and successful approach for novel material design \cite{curtarolo2013high,mounet2018two} and drug discovery \cite{kim2018new,kim2019selective}. The development of physics-based molecular dynamics (MD) simulations have enabled enormous progresses in DMI and provided nanoscale structural resolution on the insertion of drugs into lipid bilayers \cite{zhao2013engineering}. In MD simulations, the free energy profile or potential of mean force (PMF) along the membrane thickness can be extracted, which grands rich information of DMI and membrane selectivity, as shown in Figure  \ref{Fig:DMI}. To illustrate, Kim \textit{et al.} \cite{kim2019selective} applied high throughput screening for antibiotic discovery and  bithionol was discovered as a candidate. All-atom MD (AAMD) simulations were then employed to showcase the difference of the PMF profiles of bithionol across bacterial (mimicked by membranes without cholesterol) and mammalian membranes (mimicked by membranes with cholesterol). As seen from the PMF profile in Figure \ref{Fig:DMI}\textbf{a}, this drug can selectively permeate the bacterial membrane whereas avoid mammalian membrane. This HTC-screened antibiotic was validated very well by further experiments for the membrane selectivity. Their work demonstrated the capability of physics-based modeling in study of DMI problems. Though AAMD simulation is indeed a great tool to study DMI, the shortcomings associated with it are also very significant. For example, the computation time of a compound across a lipid membrane using AAMD is approximated to be $10^5$ CPU hours \cite{menichetti2017silico}. The atomic level resolution in turn severely restricts the system size that can be modeled, which thus limits the number of compounds per study only up to 10 compounds \cite{menichetti2019drug}.

To address this issue, coarse graining (CG) of all-atom system for computation speedup is widely adopted \cite{klein2008large,flores2012multiscale}. By lumping several atoms into a so-called superatom or CG bead, CG mapping can considerably reduce the degree of freedom of a biological system, and thus allows for simulation of larger systems and longer-lasting bioprocess. One noteworthy example is the Martini CG approach, which has proved very successful in modeling of biological systems \cite{marrink2007martini,monticelli2008martini}. Notably, it has been showed that using Martini CG techniques, the speedup is about $10^6$ compared to AAMD simulations, which leads to an exploration of about 500,000 molecules in study of drug-membrane permeability \cite{menichetti2019drug}. 

However, the exhaustive simulations using CGMD alone are still daunting since the combination is exponentially increased when longer beads are considered. Nonetheless, CGMD simulations have fostered big data for DMI problems in recent years \cite{hoffmann2020molecular}, which makes it very suitable for data-driven studies using advanced machine learning (ML) or deep learning (DL) algorithms. The usage of DL in drug-membrane interactions in turn enables the exploration in a much larger chemical space. Yet, an accurate and effective deep learning framework to take advantage of Martini CGMD simulations of DMI is still limited. The difficulty comes from effective feature representation of Martini drugs and membranes, as feature representation severely affects DL model performance \cite{chen2020machine}. Using SMILES string \cite{weininger1988smiles} for organic molecules from CGMD simulation losses the one-to-one mapping characteristic. It is because in coarse graining many different organic molecules can map to the same Martini CG molecules, and thereby sharing the same PMF profiles. This physically sensible rule results in poor performances of ML/DL models using SMILES as the molecular feature. For example, in a study \cite{chen2020pccp} using DNN model and SMILES as features for prediction of free energy barrier ($\Delta G$), the reported coefficient of determination or $R^2$ score was limited to 0.69. 

To address this difficulty, we develop an accurate and highly flexible deep learning framework, $\textit{i.e.}$, DMInet for DMI study. It can uniquely encode Martini CG drugs and multi-membranes, and thus is potential for high throughput screening across a vast chemical space. In addition, since it considers multi-membranes, insights are endowed by this model on membrane selectivity in design of potent drugs. Last but not least, it is highly flexible in its architecture. New CG molecules, membrane types, and/or other inputs can be easily feed into the network.

%%%%%%%%%%%%%%%%%%%%%%%%%%%%
\begin{figure*}[!htbp]
\centering
\includegraphics[width=1.0\textwidth]{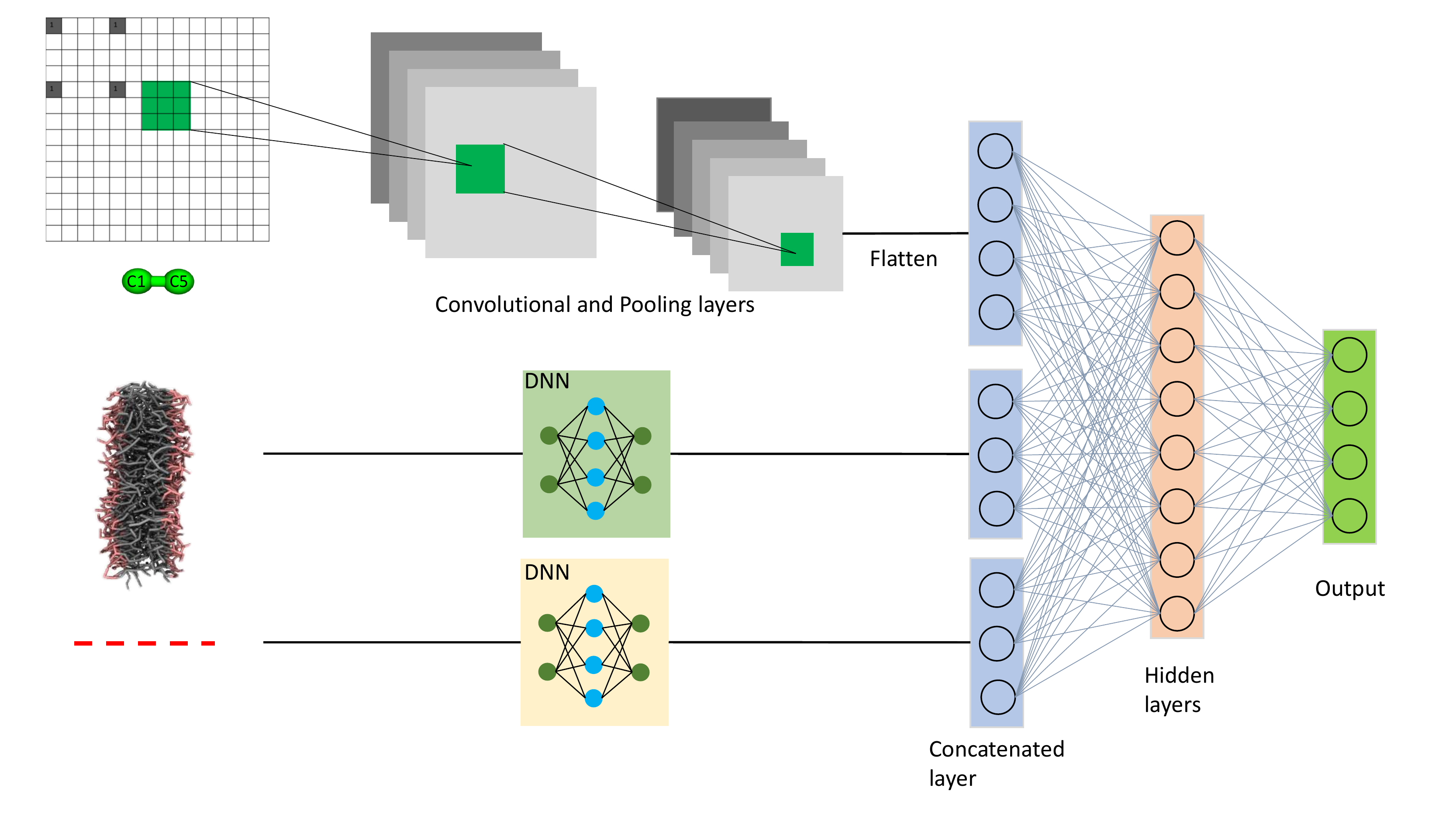}
\caption{Schematic of the DMInet architecture, which receives three inputs: the Martini CG molecule, the membrane and the spatial distance along the axis of the membrane thickness.}
\label{Fig:DMInet}
\end{figure*}

%%%%%%%%%%%%%%%%%%%%%%%%%%%%%%%%%%%%%%%%%%%%%%%%%%%%%%%%%%%%%%%%%%%%%%%%%%%%%%%%%%%%
\section{Results}
%%%%%%%%%%%%%%%%%%%%%%%%%%%%
\subsection{DMInet Architecture}
As illustrated in Figure \ref{Fig:DMInet}, the proposed DMInet model receives three distinct inputs: the image-based Martini CG molecules, the integer-labeled membrane types, and the numerical spatial distance along the membrane thickness. In general, various operations can be applied on each of them before the concatenation of the temporal outputs into a complete deep neural network (Figure \ref{Fig:DMInet}). That is, convolutional neural network (CNN) is employed for the image-based CG molecules, while deep neural network (DNN) is applied for the membrane type and the spatial distance. 

Subsequent to these neural networks, the temporary outputs are then flattened and concatenated into a dense layer which is followed by hidden layers and an ultimate output layer denoting the PMF values. In this way, the PMF profile with structural resolutions can be predicted for drug-membrane interactions and membrane selectivity.

The architecture of DMInet has high flexibility in that the layers of the CNN and DNN models in three components can be tuned for sake of accuracy as per the quantity and quality of data at hand. Namely, the number of convolutional layers in the CNN model and the number of dense layers in the two DNN models can be set accordingly. Furthermore, it can be simplified to be a combination of a single CNN model and a DNN model depending on the number of inputs of the problem being studied. The flexibility of this DL framework allows for not only drug-membrane interactions, but also wider applications to other problems involving multi-component interactions. Here, for the sake of accuracy and robustness of the model given the amount of data points in the database, only CNN-related operations are carried out.

In building the CNN part of the DMInet using the Martini CG molecules, the convolutional operation followed by max pooling and batch normalization were applied. The output of CNN-model were flattened and concatenated with the other two inputs. For detailed convolutional operations and the final model architecture, readers are referred to the Supporting Information.

%%%%%%%%%%%%%%%%%%%%%%%%%%%%
\begin{figure}
\includegraphics[width=1.0\textwidth]{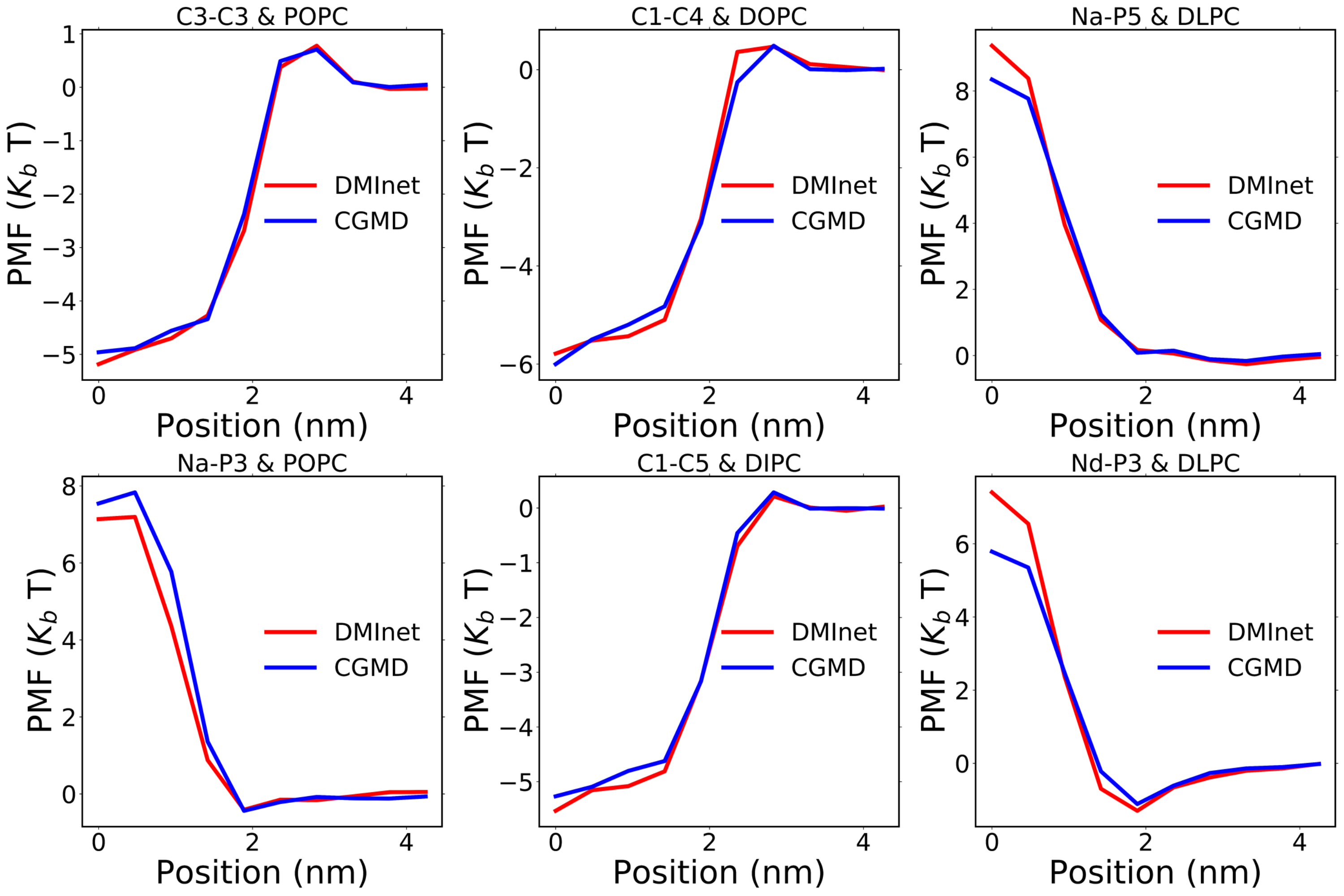}
\caption{Comparison of DMInet predictions and true PMF profiles of six unseen drug-membrane pairs.}
\label{Fig:PMF}
\end{figure}

%%%%%%%%%%%%%%%%%%%%%%%%%%%%
\subsection{Predictions of the PMF of Unseen Drug-like Molecules Across Membranes}
DMInet is trained on a training dataset with $80\%$ of the total dataset, whereas the remaining $20\%$ data are not seen by DMInet. Its prediction ability is tested on this unseen data, which contains 126 drug-membrane data points. The training hyperparameter settings and model training information can be found in the Material and Model Section and the Supporting Information.

\begin{figure}
\includegraphics[width=1.0\textwidth]{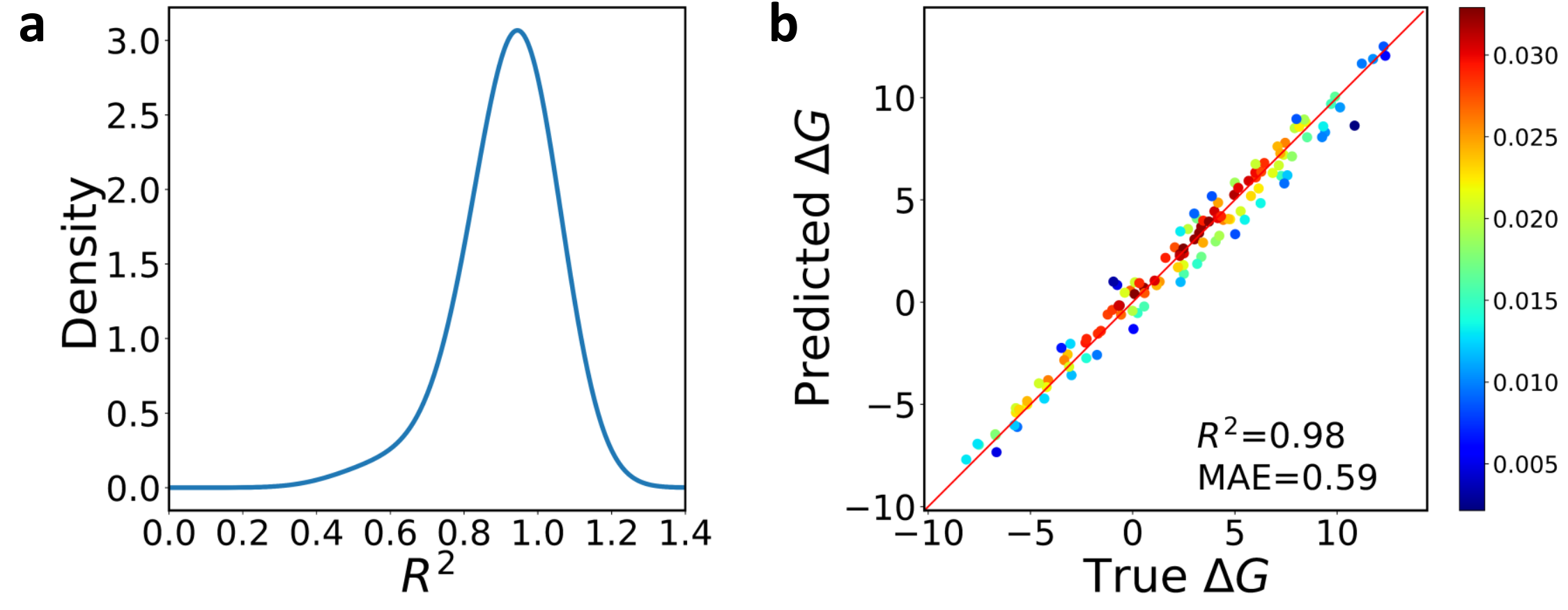}
\caption{Evaluations of DMInet on the unseen test dataset. \textbf{a}: the $R^2$ distribution of the correlations between DMInet predictions and ground truth on the 126 unseen data; \textbf{b}: the comparison of DMInet prediction and ground truth on the free energy barrier for the total dimers.}
\label{Fig:dimers}
\end{figure}

Figure \ref{Fig:PMF} gives the comparison of the PMF profile from DMInet prediction and the ground truth of six randomly selected samples. The complete comparison of all other test samples can be found in the Supporting Information. The distribution of the $R^2$ correlation score on the prediction is given in Figure \ref{Fig:dimers}\textbf{a}. One can see that DMInet can predict the whole PMF profile with reasonable accuracy. With this predicted PMF profile, insights are offered on whether a certain drug is energetically favorable or unfavorable to permeate across a lipid bilayer. Additionally, for a given drug-like molecule, the PMF profile of permeating across different membranes can grant information of membrane selectivity, which is of great practical significance in drug discovery.

Using the predicted PMF profile, the free energy barriers can be extracted of drug-like molecules across lipid membranes, which is plotted in Figure \ref{Fig:dimers}\textbf{b}. A $R^2$ score of 0.98 is obtained using DMInet, which significantly outperforms the DNN model using the SMILES string as feature representation \cite{chen2020pccp}. It demonstrates the advantages of using Martini CG as molecular representation than the SMILES strings of molecules.

%%%%%%%%%%%%%%%%%%%%%%%%%%%%

%%%%%%%%%%%%%%%%%%%%%%%%%%%%
\subsection{Predictions of the PMF of An External Trimers Data}
In addition to the interpolation test of DMInet as showed in the previous example, the extrapolation test using the trimers permeating across the DOPC membrane \cite{hoffmann2019controlled} is conducted. The objective of this test is twofold. First, it is meant to show that DMInet can accommodate more complex Martini CG molecules than dimers. Second, it is a way to test the generalization ability of DMInet for extrapolation.

\begin{figure}
\includegraphics[width=1.0\textwidth]{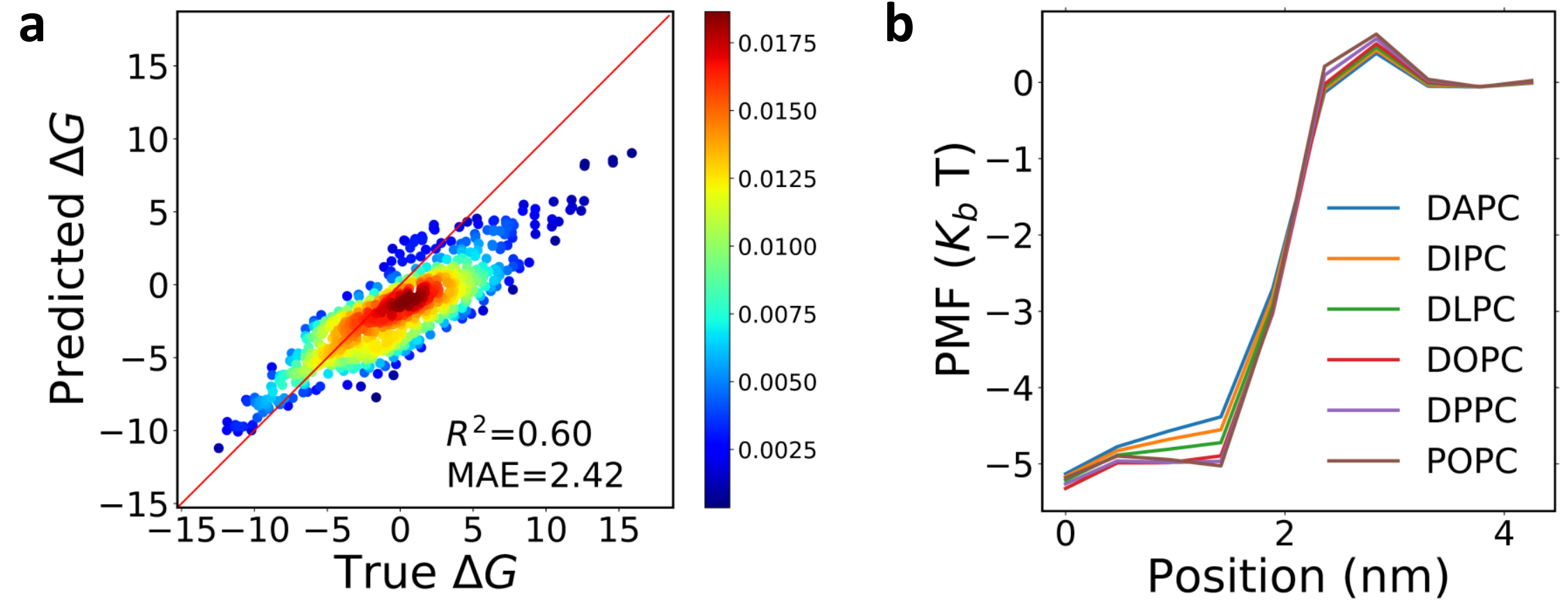}
\caption{Model prediction on a trimer dataset. \textbf{a}: the prediction of the free energy barrier for the trimers; \textbf{b}: PMF profile of Martini CG molecule (N0-C1-C4) across six membranes.}
\label{Fig:trimers}
\end{figure}

Note that in this external data \cite{hoffmann2019controlled}, only the free energy barrier values of 694 linear trimers are available rather than the whole PMF profile. DMInet can predict the whole PMF profile, which is then processed to extract the free energy barrier values. The $R^2$ correlation by DMInet predictions of this test is given in Figure \ref{Fig:trimers}\textbf{a}. One can see that the performance of this extrapolation test is worse than that of the previous example. However, the degradation of the $R^2$ and MAE in prediction is not unexpected. From the perspective of machine learning, the extrapolation ability of general ML models are not guaranteed since the extrapolated data may not fall into the same range as the training data (dimers only). This is also expected from the physical chemistry of the Martini CG molecules. For molecules with longer beads, the range of $\Delta G$ is larger than that of shorter beads \cite{chen2020machine}.

To showcase the ability of DMInet on membrane selectivity, a randomly selected Martini CG molecule (N0-C1-C4) is adopted for PMF profile prediction across six membranes. The predicted PMF profile is given in Figure \ref{Fig:trimers}\textbf{b}. As can be seen from this result, DMInet can differentiate across various membrane types. It's shown that DMInet can be a powerful tool for drug discovery with membrane selectivity.

This test shows that DMInet can allow for predictions of complex Martini CG molecules, which holds the potential for exploration in a more diversified chemical space. However, for sake of accuracy, the extrapolation test should be carefully done and interpreted. One reasonable approach, on the other hand, is to sample uniform and representative data of CG molecules, and then apply DMInet for reasonable interpolations. As a result, the computational cost can be reduced.

%%%%%%%%%%%%%%%%%%%%%%%%%%%%

%%%%%%%%%%%%%%%%%%%%%%%%%%%%%%%%%%%%%%%%%%%%%%%%%%%%%%%%%%%%%%%%%%%%%%%%%%%%%%%%%%%%
\section{Discussion}

In this work, a deep-learning framework, namely DMInet has been proposed for drug-membrane interactions with the ability to differentiate distinct membranes. DMInet uses the Martini CG beads as the molecular representation of drugs, and so its reliability and accuracy depend on the Martini CG procedure and the transferrability of the CG approach. Since the Martini CG forcefield adopts a top-down approach to reproduce the partitioning free energies \cite{marrink2007martini}, it has very good transferrability and has been widely employed for CG modeling of biomolecules such as lipids and proteins \cite{monticelli2008martini,lopez2013martini}. The efficiency of DMInet is mainly dependent on the efforts for CG mapping which can be easy or difficult as per the complexity of the molecules. 

While DMInet has demonstrated great potential in study of drug-membrane interactions in accelerated exploration of drug candidates among a larger chemical space for drug discovery, one may note there are several limitations in this work. For example, the molecules in the database is limited to Martini CG dimers and membranes are confined to only six types. Additionally, only the linear Martini CG molecules are considered while the nonlinear ones, $\textit{e.g.}$ ring molecules, are not studied. However, one should note that all these limitations are inherited either from the database (only linear Martini dimers are contained) or from the Automartini package \cite{bereau2015automated} because of the difficulty in converting complex molecules into corresponding Martini CG molecules, rather than associated with the DL framework of DMInet. In fact, DMInet is very general and has no difficulties to accommodate for complex molecules. The process would be trivial as all key points are introduced.

As CGMD simulations have been widely adopted for high throughput screening in drug discovery, more and more data will be available. Therefore, it is very beneficial to apply DL based methods to study these problems. A DL method can in turn reduces the chemical space for sampling using physic-based simulations as DL models can have exceptional interpolation capability. However, a vital point in applying DL methods to drug related problems is the molecular representation of organic molecules which heavily impacts on the performances of ML/DL models. For example, in a study \cite{chen2020machine} using DNN for $\Delta G$ prediction, the reported $R^2$ was limited to 0.69 while in this work the reported $R^2$ is as high as $0.98$. The deficiency comes from the molecular representation as using SMILES string notation for organic molecules from CGMD simulation losses the one-to-one mapping characteristic. Conversely, DMInet naturally retains the one-to-one mapping advantages by using the Martini CG molecules as features. As a result, it is superior of DMInet to make more accurate predictions.

In addition, DMInet is very flexible in that the number of inputs and ML models used for processing of each input can be varied. For example, the spatial distance (the 3rd input) can be replaced with other quantities, $\textit{e.g.}$, drug concentration or membrane thickness, to build more complex models with more diversified inputs. This flexibility is one of the advantages associated with DMInet. Moreover, it has the potential to improve itself once more computational data is available. One of the future directions the authors are focusing on is to collect computational data from cholesterol-rich membranes and normal membranes to empower DMInet for membrane selectivity problems in design of antibacterial drug.

%%%%%%%%%%%%%%%%%%%%%%%%%%%%%%%%%%%%%%%%%%%%%%%%%%%%%%%%%%%%%%%%%%%%%%%%%%%%%%%%%%%%
\section{Conclusions}
In summary, we presented a novel deep learning framework (DMInet) which has been showed effective to study drug-membrane interactions and membrane selectivity. DMInet is designed as an effective framework to take advantage of Martini CGMD simulations for accurate PMF predictions of drug-like molecules across membranes. DMInet holds the promise for accelerated drug discovery as more and more data are becoming available. We hope this framework can be helpful in drug design and serve as a basis for development of more powerful deep learning platforms to tackle more complex biochemical problems.

%%%%%%%%%%%%%%%%%%%%%%%%%%%%%%%%%%%%%%%%%%
%\section*{Conflicts of interest}
%There are no conflicts of interest to declare

\section{Material and Model}

\subsection{Database and Feature representation}
The dataset for ML model development in this work is from large scale CGMD simulations using Martini force field \cite{hoffmann2020molecular}. This dataset contains all possible dimers (CG representation of organic molecules by 2 CG beads) of  polar, apolar and nonpolar types. The enumeration of these 14 unique single beads gives rise to 105 CG dimers. In addition, the dataset also includes 6 different lipid membranes, \textit{i.e.}, DAPC (1,2-diarachidonoyl-sn-glycero-3-phosphocholine), DIPC (1,2-dilinoleoyl-sn-glycero-3-phosphocholine), DLPC (1,2-dilauroyl-sn-glycero-3-phosphocholine), DOPC (1,2-dioleoyl-sn-glycero-3-phosphocholine), DPPC (1,2-dipalmitoyl-sn-glycero-3-phosphocholine), and POPC (1-palmitoyl-2-oleoyl-sn-glycero-3-phosphocholine) membranes. The combination of 105 CG molecules and 6 membranes results in 630 unique data points of drug-membrane pairs. In this dataset, the PMF profile was obtained by umbrella sampling technique \cite{kastner2011umbrella} along the direction of membrane thickness (4.1 nm) with a step size of 0.1 nm. Therefore, the database containing CG molecules, different membranes and PMF profiles is very suitable for the purpose of this study on membrane permeability and membrane selectivity in drug-membrane interactions. 

%%%%%%%%%%%%%%%%%%%%%%%%%%%%
\begin{figure}
\includegraphics[width=1.0\textwidth]{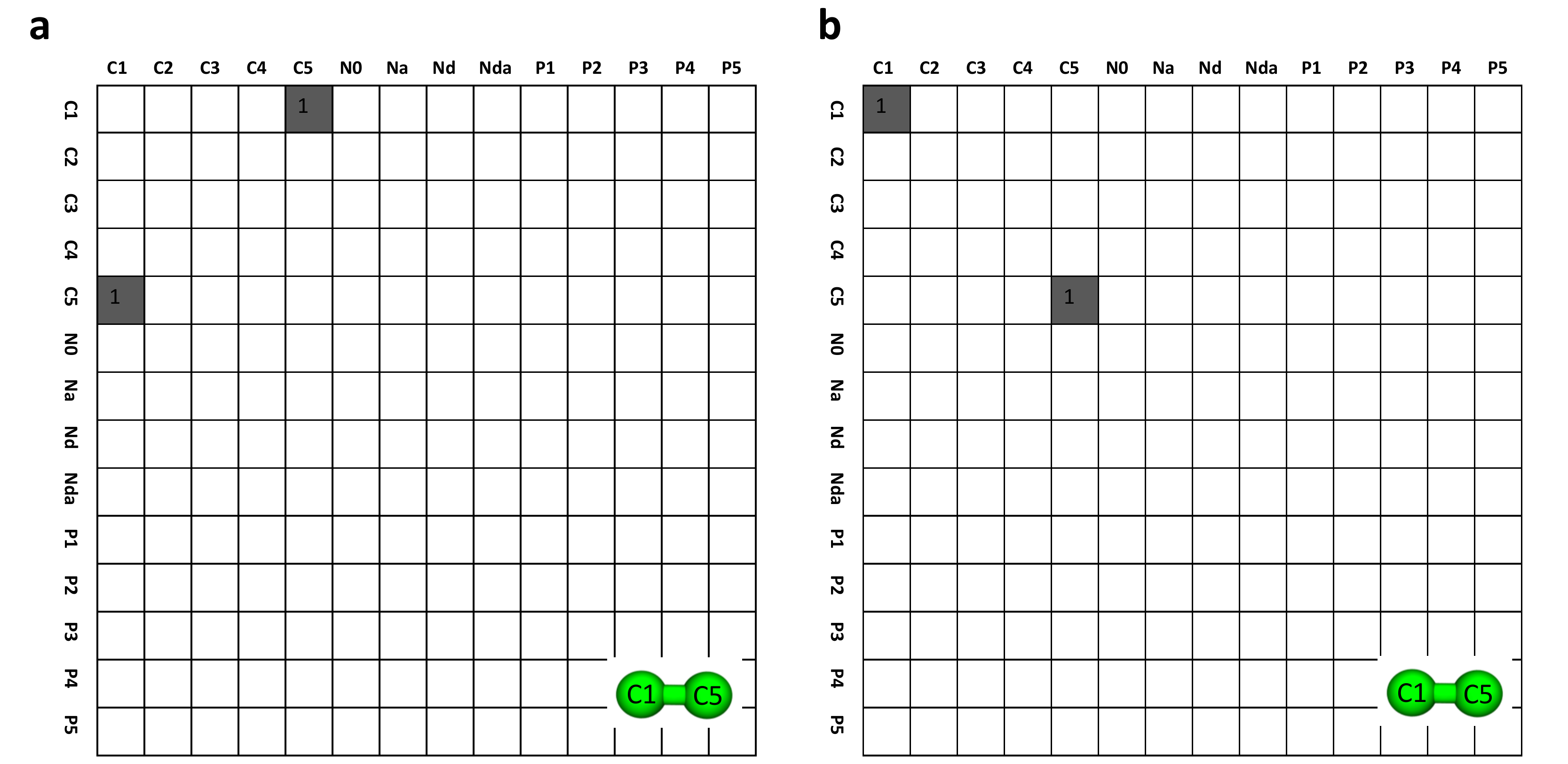}
\caption{Feature representation of CG beads. \textbf{a}: the adjacency matrix for the C1-C5 bead; \textbf{b}: the degree matrix for the C1-C5 bead.}
\label{Fig:DMIMat}
\end{figure}

To develop a ML model, the features (CG molecules and membranes in this work) should be numerically encoded so that they can be feasibly applied to ML models as inputs. For the representation of CG molecules or dimers, the idea of graph representation was employed. A square matrix of 14 in width by 14 in height is applied for Martini CG molecules, which is the summation of an adjacency matrix and a degree matrix \cite{shmilovich2020discovery} (Figure \ref{Fig:DMIMat}) so that each CG molecule can be uniquely represented. For more detailed information, readers are referred to the Supporting Information. Note that this representation can capture the symmetry of the linear Martini molecules. In addition, this representation of the CG molecule is general and can be applied for longer Martini molecules and even ring molecules.

The membrane type can be easily encoded using one-hot encoding by a vector with a length of 6. However, considering the limit of the data and ML model performance, in this work they are treated as categorical variables. Namely, an integer is assigned to each of them as DAPC=0, DIPC=1, DLPC=2, DOPC=3, DPPC=4, and POPC=5. The spatial distances are numerical numbers and can be represented by a vector.

\subsection{ML model development}
In developing the DMInet, a train/test split ratio of 0.8/0.2 was applied so that the training dataset can be used for model development and the unseen test dataset can be saved for model evaluation. In training the model using the training dataset, the checkpoints and early stopping methods were applied so that the training process can be automatically ceased when the best possible model is obtained. As a result, the overfitting possibility can be largely reduced. The model was developed under the Tensorflow platform \cite{abadi2016tensorflow} in which all aforementioned modules and techniques can be feasibly implemented. For more detailed information of the hyperparameters used and learning curves, readers are referred to the Supporting Information.

Mathematically, the DMInet seeks to find an approximation function $f:(\mathbb{R}^{d_1\times d_1}, \mathbb{R}^{d_2}, \mathbb{R}^{d_3}) \mapsto \mathbb{R}^{d}$ where $d_1$, $d_2$, and $d_3$ are respectively the dimension of the image, membrane and spatial distance vector, and $d$ is the dimension of the PMF profile. The training objective is to find the weights and biases of DMInet by minimizing the loss or cost function, which is defined as the mean squared error (MSE) of the predicted and true PMF values:
\begin{equation}
\mathcal{L}(w,b) = \frac{1}{m} \sum^{m}_{i=1} (y_i - \hat{y}_i)^2
\end{equation}
where $y_i$ and $\hat{y}_i$ are the true label and predicted values. $m$ is the number of training samples.

The evaluation metric for model performance is the $R^2$ score:
\begin{equation}
R^2 = 1 - \frac{\sum^{n}_{i=1} (y_i - \hat{y}_i)^2}{\sum^{n}_{i=1} (y_i - \frac{1}{n} \sum^{n}_{j=1} y_j)^2}
\end{equation}
and mean absolute error (MAE) of $\Delta G$:
\begin{equation}
\text{MAE} = \frac{1}{n} \sum^{n}_{i=1} |\Delta G_i - \Delta \hat{G}_i|
\end{equation}
where $n$ is the number of test samples.

\bibliography{achemso-demo}

\end{document}